\documentclass{article}

\PassOptionsToPackage{numbers, compress}{natbib}
 \usepackage[preprint]{neurips_2026}

\usepackage[utf8]{inputenc} 
\usepackage[T1]{fontenc}    
\usepackage{hyperref}       
\usepackage{url}            
\usepackage{booktabs}       
\usepackage{amsfonts}       
\usepackage{nicefrac}       
\usepackage{microtype}      
\usepackage{xcolor}         
\usepackage{amsmath}
\usepackage{amssymb}
\usepackage{bm}
\usepackage{graphicx}

\title{DuET: Note--Frame Dual Encoding for Piano Performance-to-Score Transcription}

\author{%
  Wei Zeng$^{\sharp,\flat}$
  \qquad
  Ye Wang$^{\sharp,\flat}$ \\[1.2ex]
  $^{\sharp}$Integrative Sciences and Engineering Programme, NUS Graduate School \\
  $^{\flat}$School of Computing, National University of Singapore \\[0.8ex]
  \texttt{w.zeng@u.nus.edu}
  \quad
  \texttt{dcswangy@nus.edu.sg}
}

\begin{document}

\maketitle

\begin{abstract}
Transcribing expressive piano performances into scores requires interpreting individual note timings within an underlying metrical structure. However, score attributes are associated with notes, whereas beats may occur without coincident note onsets. We introduce DuET, a \textbf{Du}al-\textbf{E}ncoder \textbf{T}ransformer that jointly learns note-level performance MIDI-to-score (PM2S) transcription and frame-level beat and downbeat tracking. Bidirectional cross-attention connects note and frame encoders, while an autoregressive decoder alternates between their representations to predict structured score attributes. We further introduce TimeRoPE and Time--Pitch Rotary Positional Encoding (TPRoPE) to establish a shared performance-time reference and explicitly encode relative temporal and pitch relationships. Evaluations on ASAP, a classical piano dataset, and our curated Gangqinpu dataset, which spans multiple genres with a focus on popular piano music, demonstrate improvements in score timing and notated duration. Compared with the strongest baseline on each dataset, DuET reduces MUSTER mean error from 6.56\% to 3.64\% on Gangqinpu and from 11.30\% to 11.22\% on ASAP. Ablations support the benefits of frame-level context, time-based positional encoding, and decoder-history masking. Code, weights, and demos are available at \url{https://github.com/wei-zeng98/DuET}.
\end{abstract}

\section{Introduction}
\label{sec:introduction}

Automatic piano transcription converts piano performances into symbolic representations. These representations are useful for music analysis, education, and downstream music information retrieval (MIR) tasks such as expressive performance rendering. Based on their inputs and outputs, transcription tasks can be grouped into audio-to-MIDI~\citep{yan2021skipping,yan2024scoring,kong2021high,hawthorne2021sequence,toyama2023automatic}, audio-to-score (A2S)~\citep{zeng2024end,carvalho2017towards,roman2019holistic,tamer2026rubato}, and performance MIDI-to-score (PM2S)~\citep{liu2022performance,beyer2024end,zeng2026bridging}. Audio-to-MIDI transcribes audio into notes with performance onset times and durations, whereas A2S and PM2S aim to produce notated scores from audio and performance MIDI, respectively. Despite substantial progress in audio-to-MIDI transcription, accurately producing a notated score remains an open problem.

The first challenge in score transcription is that mapping performance time to score time requires interpreting individual note timings within a broader rhythmic context. Under expressive timing, the same performed time interval may correspond to different score-time intervals depending on tempo and metrical context. Accurate PM2S transcription therefore requires modeling not only the temporal relationships among notes but also their alignment with the underlying metrical structure. Beat and downbeat tracking~\citep{zhao2022beat,foscarin2024beat} provides supervision for this metrical context, but its prediction granularity differs from that of note-level score transcription. Score attributes are associated with individual notes, whereas beat positions are represented on a frame-level time grid, including frames without any note onset.

A second challenge lies in representing the relative positions of notes in a musically meaningful way. Sequence-index distances do not directly reflect temporal separation: consecutive notes may occur simultaneously or be separated by a substantial time interval. Although timing encodings based on inter-onset intervals (IOIs)~\citep{beyer2024end,liu2022performance} explicitly represent adjacent-note distances, non-adjacent temporal distances require accumulating the intervening IOIs. More fundamentally, notes occupy a two-dimensional time--pitch space, as illustrated by a piano roll. Serializing these notes into a one-dimensional sequence preserves their individual attributes, but the resulting sequence order does not explicitly encode relative temporal and pitch distances.

Beyond these modeling challenges, recent neural PM2S studies have primarily relied on publicly available classical piano datasets such as ASAP~\citep{foscarin2020asap}. However, the practical need for score transcription is particularly strong for popular piano music, where corresponding sheet music is often less readily available than for classical repertoire. This motivates training and evaluation across both classical and popular piano music.

In this paper, we present a complete piano A2S pipeline comprising audio-to-performance MIDI transcription, PM2S transcription, and score formatting. We use Transkun v2~\citep{yan2024scoring} and MuseScore 4~\citep{musescore4} for the first and final stages, respectively, and focus our methodological contributions on the intermediate PM2S stage. We formulate PM2S transcription as a one-to-one mapping between input performance notes and notes in a structured score representation. The mapping preserves note identities and pitches while predicting musical attributes, including score onsets and durations, staff and voice assignments, meter and key signatures, and grace-note labels.

Under this formulation, we introduce \textbf{DuET} (\textbf{Du}al-\textbf{E}ncoder \textbf{T}ransformer), which integrates frame-level beat tracking with note-level PM2S transcription. To accommodate their different prediction granularities, DuET uses a note encoder to model individual notes and their relationships, and a frame encoder to capture temporal context on a regular time grid. Bidirectional cross-attention allows the two branches to exchange information, while alternating decoder cross-attention combines note relationships with metrical context for score prediction. This architecture enables joint learning from note-level score supervision and frame-level beat/downbeat supervision. To establish a common temporal reference across note and frame representations, we introduce \textbf{TimeRoPE}, which encodes relative performance-time distances within and across the two branches. We further propose \textbf{Time--Pitch Rotary Positional Encoding (TPRoPE)} to explicitly encode both relative temporal distances and pitch intervals in note attention.

DuET achieves the lowest MUSTER mean error among the compared methods on ASAP and Gangqinpu, as well as the lowest onset, offset, duration, and staff-assignment errors on both datasets. Ablation studies examine the contributions of dual encoding, positional encoding, decoder memory access, and masking. Our main contributions are:
\begin{itemize}
    \item We introduce \textbf{DuET}, a note--frame dual-encoder Transformer that jointly learns frame-level beat and downbeat tracking and note-level PM2S transcription through bidirectional encoder cross-attention and alternating decoder cross-attention.
    \item We introduce a shared performance-time coordinate system through \textbf{TimeRoPE} and propose \textbf{TPRoPE} to explicitly encode relative temporal and pitch relationships in note attention.
    \item We evaluate DuET on classical piano performances from ASAP and our curated Gangqinpu dataset, which spans multiple styles with a focus on popular piano music. DuET achieves the lowest MUSTER mean error among the compared methods on both datasets. We provide inference code, pretrained weights, and an interactive audio-to-score demo.
\end{itemize}

\section{Related work}
\label{sec:related_work}

\subsection{Piano performance MIDI-to-score transcription}
Early PM2S methods used probabilistic models to estimate rhythmic structure and notation attributes~\citep{raphael2001automated,cogliati2016transcribing,shibata2021non}. Neural approaches subsequently introduced recurrent networks for joint note-value and voice estimation~\citep{hiramatsu2021joint} and beat-based rhythm quantization~\citep{liu2022performance}. More recently, Transformer encoder--decoder models have formulated PM2S transcription as sequence-to-sequence prediction of structured scores~\citep{beyer2024end,wachter2025transformer,zeng2026bridging,ye2026magic}.

Within this progression, metrical information has been incorporated in different ways. \citet{liu2022performance} explicitly predict beat and downbeat labels at note onsets and use dynamic programming to infer beats without coincident note onsets. This additional inference addresses the mismatch between note-based predictions and a metrical grid that extends beyond note onsets. Several Transformer encoder--decoder models infer metrical structure implicitly through note-associated score positions and measure information~\citep{beyer2024end,zeng2026bridging}. \citet{wachter2025transformer} take another approach, conditioning rhythm quantization on an externally supplied beat and downbeat grid. When this grid is unavailable, a separate beat tracker is required, and its errors can propagate through beat-based preprocessing into quantization. DuET jointly learns frame-level beat tracking and note-level score prediction, providing direct supervision at both granularities without requiring an externally supplied beat grid at inference.

Beyond metrical modeling, repertoire coverage remains limited. Prior work has largely evaluated PM2S transcription on classical piano datasets, such as ASAP~\citep{foscarin2020asap}, while \citet{shibata2021non} also examined popular piano music. Although \citet{wachter2025transformer} additionally explored jazz guitar, their piano evaluation remained focused on classical repertoire. We complement these studies by training and evaluating DuET on both classical and popular piano music.

\subsection{Positional encoding in symbolic music}

Relative positional information is important for modeling relationships in symbolic music. Music Transformer~\citep{huang2019music} introduced efficient relative attention to capture dependencies over long musical sequences. RoFormer~\citep{su2024roformer} subsequently introduced rotary positional encoding (RoPE), which applies position-dependent rotations to queries and keys so that their attention scores depend on relative positions. \citet{beyer2024end} adopted RoPE in a Transformer encoder--decoder for PM2S transcription, combining sequence-position encoding with compound note representations.

These positional encodings primarily describe sequence-index relationships, rather than explicitly modeling both performance-time distances and pitch intervals. In computer vision, two-dimensional RoPE has been investigated to encode spatial relationships~\citep{heo2024rotary}. We adapt this multidimensional principle to symbolic music through TPRoPE, using time and pitch coordinates for note attention and a shared performance-time reference across note and frame representations.

\section{Methodology}

\subsection{Task formulation}
\label{sec:task_formulation}

We study PM2S transcription within a three-stage A2S pipeline. We use Transkun v2~\citep{yan2024scoring} to obtain performance MIDI from audio and MuseScore 4~\citep{musescore4} to produce the final notation in MusicXML format. Our model addresses the intermediate PM2S stage, predicting the musical attributes of a structured score representation. Additionally, we jointly learn frame-level beat and downbeat tracking as an auxiliary task to support metrical understanding. Notation and engraving decisions, including pitch spelling and beaming, are delegated to MuseScore 4.

\paragraph{Performance MIDI input.}
For an input window, we represent the performance as a sequence of $N$ notes:
\begin{equation}
    \mathbf{X}=(x_1,\ldots,x_N),
    \qquad x_i=(p_i,t_i,d_i,v_i),
\end{equation}
where $p_i\in\{0,\ldots,127\}$ is the MIDI pitch, $t_i$ and $d_i$ are the onset and duration in seconds, and $v_i\in\{1,\ldots,127\}$ is the note-on velocity. Notes are ordered by performance onset, retaining their original input order for simultaneous notes. From the same note sequence and available sustain-pedal messages, we derive a piano-roll representation $\mathbf{R}=(r_1,\ldots,r_T)$, where $T$ is the number of frames. Frames are sampled at times $t_j^{\mathrm{frame}}=t_{\mathrm{start}}+(j-1)\Delta$, with $\Delta=0.02$ seconds and $t_{\mathrm{start}}$ denoting the start of the input window. Each $r_j$ describes pitch activity and pedal state at that time.

\paragraph{Note-aligned score output.}
We formulate PM2S transcription as a one-to-one correspondence between the input notes $\mathbf{X}$ and output score records $\mathbf{Y}=(y_1,\ldots,y_N)$, preserving the input note order. Each output record
\begin{equation}
    y_i=\bigl(p_i,b_i,q_i,f_i,\ell_i,s_i,L_i,m_i^{\mathrm{num}},m_i^{\mathrm{den}},k_i,u_i,g_i\bigr)
\end{equation}
contains the score attributes defined in \autoref{tab:score_attributes}. Pitch is copied from the corresponding input note, while the remaining attributes are predicted by the model. Bar-level attributes, including bar length, time signature, and key signature, are predicted within each note record rather than as separate bar tokens. The bar index $b_i$ is encoded relative to the window's bar anchor $a$. The discrete vocabularies $\mathcal{F}$, $\mathcal{D}$, and $\mathcal{L}$ are derived from onset fractions, note durations, and bar lengths observed in the training data. For $k_i$, positive values count sharps, negative values count flats, and zero indicates neither.

The one-to-one mapping excludes note insertion, deletion, and many-to-one ornament reduction during prediction. Subsequent score formatting may split a note into tied notation.

\begin{table}[t]
    \centering
    \small
    \setlength{\tabcolsep}{4pt}
    \renewcommand{\arraystretch}{1.12}
    \caption{Attributes in each note-aligned score record.
    Score-time values are measured in quarter-note units.}
    \label{tab:score_attributes}
    \begin{tabular*}{\linewidth}{@{}l@{\extracolsep{\fill}}lll@{}}
        \hline
        Attribute & Name & Values & Description \\
        \hline
        $p_i$ & Pitch & $\{0,\ldots,127\}$ & MIDI pitch copied from the input note \\
        $b_i$ & Bar index & $b_i-a\in\{-256,\ldots,255\}$ & Encoded relative to window anchor $a$ \\

        $q_i$ & Onset integer part &
        $\{0,\ldots,63\}$ &
        Whole quarter-note units from bar start \\

        $f_i$ & Onset fractional part &
        $\mathcal{F}\subset[0,1)$ &
        Fractional quarter-note offset \\

        $\ell_i$ & Duration &
        $\mathcal{D}\subset[0,51]$ &
        Notated duration; zero denotes a grace note \\

        $s_i$ & Staff &
        $\{0,1\}$ &
        Upper or lower staff, respectively \\

        $L_i$ & Bar length &
        $\mathcal{L}\subset[0,36]$ &
        Duration of the containing bar \\

        $m_i^{\mathrm{num}}$ & Meter numerator &
        $\{1,\ldots,64\}$ &
        Upper number of the time signature \\

        $m_i^{\mathrm{den}}$ & Meter denominator &
        $\{1,2,4,8,16,32,64\}$ &
        Note-value denominator of the time signature \\

        $k_i$ & Key signature &
        $\{-7,\ldots,7\}$ &
        Signed number of sharps \\

        $u_i$ & Voice &
        $\{0,\ldots,15\}$ &
        Categorical voice identifier within the window \\

        $g_i$ & Grace indicator &
        $\{0,1\}$ &
        Ordinary or grace note, respectively \\
        \hline
    \end{tabular*}
\end{table}

\paragraph{Frame-level beat output.}
We jointly learn beat and downbeat tracking on the same time grid as $\mathbf{R}$. Let $\mathbf{B}=(B_1,\ldots,B_T)$ denote the metrical labels, where $B_j\in\{0,1,2\}$ indicates a non-beat frame, an ordinary beat, or a downbeat, respectively. The frame branch predicts a three-class probability vector $\boldsymbol{\pi}_j$, with
\begin{equation}
    \pi_{j,c}=P_{\theta}(B_j=c\mid\mathbf{X},\mathbf{R}),\qquad c\in\{0,1,2\}.
\end{equation}
These auxiliary targets support metrical modeling and are separate from the output score records.

\paragraph{Autoregressive score modeling.}
We model the aligned mapping using an encoder with an autoregressive decoder:
\begin{equation}
    P_{\theta}(\mathbf{Y}\mid\mathbf{X},\mathbf{R})
    =
    \prod_{i=1}^{N}
    P_{\theta}(y_i\mid y_{<i},\mathbf{X},\mathbf{R}).
\end{equation}
At decoding step $i$, the decoder predicts the score attributes of input note $x_i$ conditioned on the encoded performance and preceding score records; its pitch is copied from the input. This autoregressive formulation models dependencies among score records without changing their predefined correspondence to input notes. The output length is fixed to $N$, so neither end-of-sequence prediction nor learned note alignment is required.

\subsection{DuET architecture}
\label{sec:model_overview}

\begin{figure}[t]
    \centering
    \includegraphics[width=\textwidth]{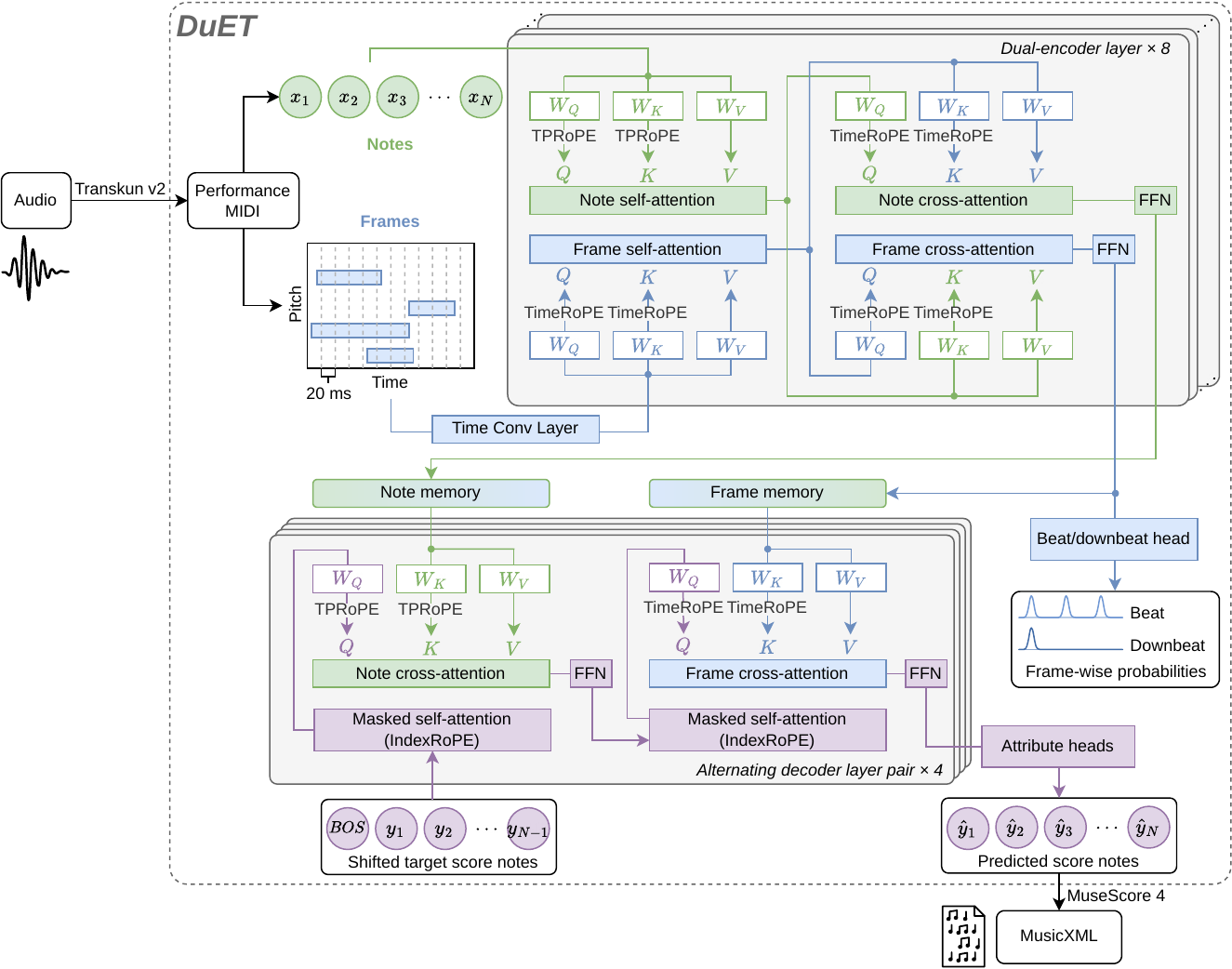}
    \caption{Overview of DuET within the audio-to-score pipeline.
    Each dual-encoder layer combines branch-specific self-attention,
    bidirectional cross-attention, and feed-forward networks.
    Successive decoder layers attend to note and frame memories in
    alternation, predicting one score record per input note.
    Rotary positional encoding is applied to queries and keys.
    Residual connections and layer normalization are omitted for clarity.}
    \label{fig:architecture}
    \vspace{-10pt}
\end{figure}

As illustrated in \autoref{fig:architecture}, DuET consists of a note encoder, a frame encoder, and a note-aligned autoregressive decoder. The two encoders accommodate different prediction granularities: score attributes are associated with individual notes, whereas beat tracking requires temporal coverage beyond note onsets. Bidirectional cross-attention connects the encoder branches, and alternating decoder cross-attention combines their output memories for score prediction. An auxiliary beat/downbeat head supervises the final frame representations.

\paragraph{Input embeddings.}
DuET uses separate embedding modules for the note sequence $\mathbf{X}$ and piano-roll representation $\mathbf{R}$ defined in Section~\ref{sec:task_formulation}, producing representations of the same hidden dimension. For each note, a learned pitch embedding is concatenated with a linear projection of its continuous performed duration in seconds and velocity normalized to $[0,1]$ by division by $127$. Performance onset is represented through positional encoding rather than note embedding to model relative timing independently of the absolute time origin. Unlike IOIs, which explicitly encode only adjacent-note intervals, this encoding directly represents temporal distances between any pair of notes. Pitch is included in both note embedding and positional encoding to capture absolute pitch identity and relative pitch intervals, respectively. The positional encoding is detailed in Section~\ref{sec:tprope}.

For the frame branch, each frame of $\mathbf{R}$ contains four pitch-wise features over the 88 piano keys (MIDI pitches 21--108): binary key-held activity, pedal-extended activity, binary note onsets, and onset velocities normalized by 127. Two additional features encode the sustain-pedal state and the availability of pedal information, yielding a 354-dimensional input per frame. Two temporal convolutional layers with GELU activations map $\mathbf{R}$ into frame embeddings. Both layers use a kernel size of $5$, stride $1$, and padding $2$, preserving the 20\,ms frame interval without temporal downsampling.

\paragraph{Dual encoder.}
The encoder comprises eight coupled layers, each with separate note and frame branches. Each layer first applies self-attention independently within the two branches, followed by bidirectional cross-attention. In note cross-attention, queries come from the note branch and keys and values from the frame branch; frame cross-attention reverses these roles. Both cross-attention modules use the note and frame states produced by self-attention, without using each other's updated outputs. A separate SwiGLU feed-forward network completes each branch, with pre-layer normalization and residual connections applied around each attention and feed-forward module. The final states are normalized at the encoder output and further normalized to form note memory $\mathbf{M}^{N}\in\mathbb{R}^{N\times d}$ and frame memory $\mathbf{M}^{F}\in\mathbb{R}^{T\times d}$ for the decoder, where $N$ and $T$ are the numbers of notes and frames, respectively, and $d$ is the hidden dimension. This interaction allows note representations to incorporate temporal context and frame representations to incorporate note relationships.

\paragraph{Frame-level objectives.}
We attach prediction heads to the final frame-encoder representations, before the additional normalization used to construct decoder memory $\mathbf{M}_F$. These heads predict beat/downbeat classes, within-bar beat indices, and the cosine and sine components of bar phase, providing three complementary supervision signals:
\begin{equation}
    \mathcal{L}_{\mathrm{frame}}
    =
    \mathcal{L}_{\mathrm{beat}}
    + 0.5\,\mathcal{L}_{\mathrm{index}}
    + 0.1\,\mathcal{L}_{\mathrm{frame\text{-}phase}}.
\end{equation}
Here, $\mathcal{L}_{\mathrm{beat}}$ is a weighted three-class cross-entropy loss for non-beat, ordinary beat, and downbeat predictions, using Gaussian soft targets with a standard deviation of one frame and class weights $(1,5,5)$. $\mathcal{L}_{\mathrm{index}}$ is a soft-target cross-entropy loss for the beat index within a bar, supervised around reliable beat annotations. $\mathcal{L}_{\mathrm{frame\text{-}phase}}$ is the squared error between predicted and target cosine and sine components of normalized bar phase, with targets interpolated between adjacent reliable beats. Each loss is normalized by its effective supervision weight, excluding padded frames and unreliable annotations.

\paragraph{Note-aligned decoder.}
The decoder comprises eight layers arranged as four successive layer pairs. Each layer applies causal self-attention, cross-attention, and a SwiGLU feed-forward network, with pre-layer normalization and residual connections. Cross-attention alternates between note memory $\mathbf{M}^{N}$ in odd-numbered layers and frame memory $\mathbf{M}^{F}$ in even-numbered layers, allowing each layer pair to combine note relationships with temporal context. Decoder self-attention uses IndexRoPE, while cross-attention uses TPRoPE for note memory and TimeRoPE for frame memory, with query coordinates taken from the corresponding input note (see Section~\ref{sec:tprope}). Separate attribute heads map the final normalized decoder state to categorical distributions for the predicted score fields, producing them in parallel at each note step. An additional pitch head provides auxiliary supervision during training.

\paragraph{Note-level objectives.}
The decoder is supervised by categorical score-attribute prediction, auxiliary pitch prediction, and note-phase consistency:
\begin{equation}
    \mathcal{L}_{\mathrm{note}}
    =
    \sum_{a\in\mathcal{A}}\lambda_a\mathcal{L}_a
    + \mathcal{L}_{\mathrm{pitch}}
    + 0.1\,\mathcal{L}_{\mathrm{note\text{-}phase}},
\end{equation}
where $\mathcal{A}$ contains the eleven predicted score fields in \autoref{tab:score_attributes}, excluding the copied pitch. Each $\mathcal{L}_a$ is a cross-entropy loss, with $\lambda_a=0.2$ for key signature, voice, and grace-note status, and $\lambda_a=1$ otherwise. The auxiliary loss $\mathcal{L}_{\mathrm{pitch}}$ supervises input-pitch prediction from the decoder state. The note-phase loss measures expected circular onset error within the reference bar using the predicted onset-integer and onset-fraction distributions, and is applied only to non-grace notes with valid timing annotations. Each loss is normalized over its valid supervised notes; padding, conditioning-prefix records, and unknown labels are excluded.

The overall training objective combines note-level and frame-level supervision:
\begin{equation}
    \mathcal{L}=\mathcal{L}_{\mathrm{note}}+\mathcal{L}_{\mathrm{frame}}.
\end{equation}

\subsection{TimeRoPE and TPRoPE}
\label{sec:tprope}

\begin{figure}[t]
    \centering
    \includegraphics[width=\linewidth]{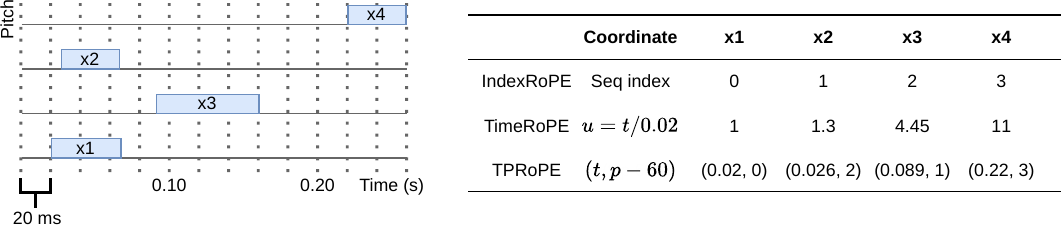}
    \caption{Positional coordinates for the same four notes. IndexRoPE uses sequence indices, TimeRoPE uses scaled performance timestamps, and TPRoPE uses both performance timestamps and pitch coordinates. The dashed vertical lines indicate the 20\,ms frame grid; note onsets need not coincide with this grid.}
    \label{fig:tprope}
\end{figure}

DuET assigns positional coordinates according to the information represented by each attention module. \autoref{fig:tprope} illustrates how the same notes receive different coordinates under IndexRoPE, TimeRoPE, and TPRoPE.

\paragraph{TimeRoPE.}
TimeRoPE replaces frame and note indices with a shared performance-time coordinate $u=(t-t_{\mathrm{start}})/\Delta$, where $t_{\mathrm{start}}$ is the input-window start and $\Delta=0.02$ seconds. Dividing by $\Delta$ expresses elapsed time in units of the 20\,ms frame interval. For frames, $t$ lies on this regular grid; for notes, it is the continuous performance onset, without rounding or quantization. For example, the onset of $x_2$ in \autoref{fig:tprope} is $0.026$ seconds, giving a non-integer coordinate of $1.3$. Applying rotary encoding to projected queries and keys makes the positional contribution to attention depend on $u_j-u_i=(t_j-t_i)/\Delta$. This shared coordinate system represents elapsed time within and across the two branches. Values remain unchanged, as in standard RoPE~\citep{su2024roformer}. Since each frame represents activity across multiple pitches and has no unique pitch coordinate, TimeRoPE is used in frame self-attention, both directions of encoder cross-attention, and decoder cross-attention to frame memory.

\paragraph{TPRoPE.}
Unlike frames, each note has a unique pitch coordinate in addition to its performance onset. TPRoPE therefore encodes both temporal and pitch relationships within every attention head. For a head dimension $d_h$, the first $d_h/2$ components of each projected query and key form the time subspace, and the remaining $d_h/2$ components form the pitch subspace. The coordinates are continuous window-relative time $t_i-t_{\mathrm{start}}$ in seconds and centered pitch $p_i-60$ in semitones. Unlike TimeRoPE, time is not divided by $\Delta$, because the temporal rotary periods are specified directly in seconds. The two subspaces are rotated independently:
\begin{equation}
    \widetilde{\mathbf{q}}_i=
    \left[
        \mathcal{R}_{t}(t_i-t_{\mathrm{start}})\mathbf{q}_i^{t};
        \mathcal{R}_{p}(p_i-60)\mathbf{q}_i^{p}
    \right],
\end{equation}
where $[\cdot;\cdot]$ denotes concatenation and $\mathbf{q}_i^{t},\mathbf{q}_i^{p}\in\mathbb{R}^{d_h/2}$ are the time and pitch components of the projected query. Each subspace contains $d_h/4$ rotary pairs, with rotation angle $2\pi x/P$ for coordinate $x$ and period $P$. The periods are geometrically spaced from $0.1$ to $64$ seconds for time and from $12$ to $256$ semitones for pitch. Keys undergo the same transformation, making the positional contribution to attention depend on relative performance-time distances $t_j-t_i$ and pitch intervals $p_j-p_i$. Values remain unchanged. TPRoPE is used in note self-attention and decoder cross-attention to note memory, where each decoder query uses the onset and pitch of its corresponding input note.

\paragraph{IndexRoPE.}
Decoder self-attention uses standard sequence-index RoPE, termed IndexRoPE, to represent score-record order. Unlike the performance coordinates used in encoder and cross-attention modules, these indices identify autoregressive decoding positions. A causal mask prevents attention to future decoder positions.

\subsection{Windowing and continuation}
\label{sec:windowing}

During training and inference, we process performances using windows containing preceding context and new notes to transcribe. The new region spans $[s,s+W)$ in performance time, where $s$ is its start time and $W$ its duration in seconds, and includes notes whose onsets fall within this interval. We use $W=32$ seconds. The preceding context includes performance inputs and a conditioning prefix of score records from up to four preceding bars. The encoders process both context and new inputs, while the decoder conditions on the score prefix to predict new records. Training uses reference score records as the prefix, with note-level losses applied only to new records.

At inference, windows advance through non-overlapping new regions. Each window takes its prefix from the most recent four bars of the accumulated predictions, together with the corresponding performance context. The first window has no score prefix. Prefix records remain unchanged, and only new predictions are appended, ensuring that each input note is transcribed once.

\subsection{Score reconstruction}
\label{sec:score_reconstruction}

After processing all windows, we reconstruct a full score from the accumulated note records. Relative bar indices are converted to full-piece indices using each window's anchor. Each note predicts a bar length, time signature, and key signature. For each bar, we discard candidate tuples whose bar length does not accommodate all predicted within-bar note onsets, and select the most frequent remaining tuple. Empty bars inherit the preceding time and key signatures, with their lengths determined by the inherited meter.

Starting from score time zero, bar lengths are accumulated to obtain the bar starts $A_b$. Each note is then placed at $\tau_i=A_{b_i}+q_i+f_i$, retaining its input pitch and predicted duration, staff, voice, and grace-note status. The resulting structured score is converted into notation using MuseScore 4, which handles pitch spelling, accidental display, stem directions, beaming, and tie reconstruction.

\section{Experimental setup}

\paragraph{Datasets.}
We use (n)ASAP~\citep{peter2023automatic}, the note-aligned extension of ASAP~\citep{foscarin2020asap}, and our curated Gangqinpu dataset\footnote{We plan to publicly release Gangqinpu and document its construction in a forthcoming dataset paper.} for training and evaluation, with additional training data from PERiScoPe~\citep{borovik2025symupe}.
(n)ASAP provides expressive piano performances paired with symbolic scores, note alignments, and beat annotations, focusing on Western classical repertoire.
Gangqinpu contains 18,787 performance--score pairs, spanning popular, classical, and other musical styles, with performance MIDI transcribed from audio using Transkun v2.
We additionally use PERiScoPe data imported through PianoCoRe~\citep{borovik2026pianocore}; its web-collected performances were also transcribed using Transkun v2, reducing differences attributable to the choice of transcription system.
We prepare note correspondences and metrical supervision to support joint score prediction and beat tracking.
For Gangqinpu and (n)ASAP, training, validation, and test splits are defined at the piece level, using cross-dataset work groups to prevent overlap.

\paragraph{Masking.}
\label{sec:masking}
During training, we apply bar-level masking to the performance-note input with probability 0.1, replacing the content embeddings of all notes within each selected bar with a learned mask embedding.
The corresponding current-note pitch embeddings supplied to the decoder are also masked, while the original onset and pitch coordinates used by RoPE are retained.
Independently, we mask each shifted decoder-history note embedding with probability 0.75 using a separate learned mask embedding, excluding BOS and padding positions~\citep{beyer2024end}.
Prediction targets remain unchanged, and both masking mechanisms are disabled during evaluation and inference.

\paragraph{Training configuration.}
We use a hidden dimension of $512$, eight attention heads, and a SwiGLU~\citep{shazeer2020glu} intermediate dimension of $1408$. DuET is trained from random initialization for $50$ epochs using AdamW~\citep{loshchilov2019decoupled} with a peak learning rate of $2\times10^{-4}$ and weight decay of $0.01$. Each epoch samples $20{,}000$ windows. Training uses two GPUs with $16$ windows per GPU, giving a global batch size of $32$ without gradient accumulation. The learning rate follows a cosine decay schedule with linear warmup over the first $5\%$ of training steps. Gradients are clipped to a norm of $1$.


\section{Results}
\label{sec:results}


\subsection{PM2S transcription performance}
\label{sec:pm2s_results}

We report PM2S transcription results on 59 performances from the ASAP test split and common-valid subsets of 79 performances from the Gangqinpu test split. We compare DuET with MuseScore, the J-pop and classical configurations of \citet{shibata2021non}, neural beat tracking~\citep{liu2022performance}, and the Transformer model of \citet{beyer2024end}, using MUSTER~\citep{nakamura2018towards,hiramatsu2021joint} and ScoreSimilarity~\citep{suzuki2021score,cogliati2017metric}. ASAP baseline results are taken from \citet{beyer2024end}. For quantitative evaluation, DuET predictions are serialized directly to MusicXML without MuseScore post-processing; MuseScore 4 is used for visualization.

As shown in \autoref{tab:pm2s_results}, DuET achieves the lowest MUSTER mean error on the Gangqinpu test split, reducing it from $6.56\%$ for the strongest baseline to $3.64\%$. It also achieves the lowest errors across all four reported ScoreSimilarity metrics. On ASAP, DuET obtains a MUSTER mean error of $11.22\%$, compared with the published Transformer result of $11.30\%$, with lower onset, offset, duration, and staff-assignment errors. Its missing-note and extra-note errors remain higher than those of several baselines. Across both datasets, the most consistent gains lie in score timing and notated duration.
\begin{table}[t]
    \centering
    \scriptsize
    \setlength{\tabcolsep}{2pt}
    \renewcommand{\arraystretch}{1.0}
    \setlength{\aboverulesep}{1.5pt}
    \setlength{\belowrulesep}{1.5pt}
    \caption{PM2S transcription errors (\%). Lower is better. Bold and underlined values indicate the best and second-best results within each dataset, ranked before rounding where full-precision results are available. ASAP baseline results are taken from \citet{beyer2024end}.}
    \label{tab:pm2s_results}
    \begin{tabular*}{\linewidth}{@{}l@{\extracolsep{\fill}}rrrrrrrrrr@{}}
        \toprule
        & \multicolumn{6}{c}{MUSTER~\citep{nakamura2018towards,hiramatsu2021joint} }
        & \multicolumn{4}{c}{ScoreSimilarity~\citep{suzuki2021score,cogliati2017metric}} \\
        \cmidrule(lr){2-7}
        \cmidrule(lr){8-11}
        Method & Pitch & Miss & Extra & Onset & Offset & Mean
        & Miss & Extra & Duration & Staff \\
        \midrule
        \multicolumn{11}{c}{\textbf{ASAP}} \\
        MuseScore~\citep{musescore4}
        & 2.41 & 7.35 & 9.64 & 47.90 & 49.44 & 23.35
        & 16.17 & 16.74 & 55.23 & 21.87 \\
        HMMs (J-pop)~\citep{shibata2021non}
        & \underline{2.09} & \textbf{6.38} & \underline{8.67} & 25.02 & 29.21 & 14.27
        & \underline{10.80} & 11.39 & 71.38 & -- \\
        HMMs (classical)~~\citep{shibata2021non}
        & 2.11 & \underline{6.47} & 8.75 & 22.58 & 29.84 & 13.95
        & \textbf{10.74} & \underline{11.28} & 64.73 & -- \\
        Neural beat tracking~\citep{liu2022performance}
        & \textbf{2.02} & 6.81 & 9.01 & 68.28 & 54.11 & 28.04
        & 17.10 & 17.67 & 66.98 & 6.86 \\
        Beyer and Dai~\citep{beyer2024end}
        & 3.11 & 7.56 & \textbf{6.44} & \underline{15.55} & \underline{23.84} & \underline{11.30}
        & 12.69 & \textbf{9.06} & \underline{51.86} & \underline{6.62} \\
        \textbf{DuET (ours)}
        & 2.69 & 8.78 & 11.70 & \textbf{12.34} & \textbf{20.59} & \textbf{11.22}
        & 13.16 & 13.70 & \textbf{36.46} & \textbf{4.98} \\
        \midrule
        \multicolumn{11}{c}{\textbf{Gangqinpu}} \\
        MuseScore~\citep{musescore4}
        & 0.81 & 3.45 & 3.26 & 16.09 & 38.80 & 12.48
        & 6.03 & 5.63 & 58.90 & 17.98 \\
        HMMs (J-pop)~\citep{shibata2021non}
        & \underline{0.65} & \underline{1.78} & \underline{2.19} & 5.75 & \underline{22.43} & 6.56
        & 3.08 & 3.18 & \underline{45.36} & -- \\
        HMMs (classical)~\citep{shibata2021non}
        & 0.65 & 1.87 & 2.28 & \underline{5.11} & 22.88 & \underline{6.56}
        & \underline{2.94} & 3.04 & 47.40 & -- \\
        Neural beat tracking~\citep{liu2022performance}
        & \textbf{0.63} & \textbf{1.71} & \textbf{1.50} & 15.33 & 40.70 & 11.98
        & 3.39 & 2.98 & 63.02 & 6.77 \\
        Beyer and Dai~\citep{beyer2024end}
        & 0.81 & 3.12 & 3.00 & 11.30 & 25.02 & 8.65
        & 4.26 & \underline{2.94} & 70.62 & \underline{4.98} \\
        \textbf{DuET (ours)}
        & 0.68 & 2.08 & 2.87 & \textbf{2.54} & \textbf{10.01} & \textbf{3.64}
        & \textbf{2.77} & \textbf{2.37} & \textbf{36.22} & \textbf{2.80} \\
        \bottomrule
    \end{tabular*}
\end{table}

\begin{table}[t]
    \centering
    \scriptsize
    \setlength{\tabcolsep}{2pt}
    \renewcommand{\arraystretch}{1.0}
    \setlength{\aboverulesep}{1.5pt}
    \setlength{\belowrulesep}{1.5pt}
    \caption{PM2S ablations on fixed common-valid subsets of 39 ASAP and
    72 Gangqinpu performances. Errors are percentages; lower is better.
    Bold and underlined values indicate the best and second-best results
    within each dataset, ranked before rounding.
    Positional/memory variants are named as ``RoPE / decoder memory''.
    DuET uses Time--Pitch RoPE with alternating memory access.
    Thin rules separate encoder, positional/memory, and masking controls.}
    \label{tab:pm2s_ablations}
    \begin{tabular*}{\linewidth}{@{}l@{\extracolsep{\fill}}rrrrrrrrrr@{}}
    \toprule
    & \multicolumn{6}{c}{MUSTER~\citep{nakamura2018towards,hiramatsu2021joint}}
    & \multicolumn{4}{c}{ScoreSimilarity~\citep{suzuki2021score,cogliati2017metric}} \\
    \cmidrule(lr){2-7}\cmidrule(lr){8-11}
    Model & Pitch & Miss & Extra & Onset & Offset & Mean
          & Miss & Extra & Duration & Staff \\
    \midrule
    \multicolumn{11}{c}{\textbf{ASAP}} \\
    DuET
    & \textbf{2.25} & \underline{5.95} & \underline{11.10}
    & \textbf{7.66} & \textbf{16.42} & \textbf{8.68}
    & \textbf{9.28} & \textbf{12.69} & 30.58 & \textbf{3.99} \\
    Note encoder only
    & 2.35 & \textbf{5.94} & \textbf{10.94}
    & 10.85 & 22.42 & 10.50
    & \underline{9.71} & \underline{13.11} & 52.23 & 4.74 \\
    \cmidrule(lr){1-11}
    Index / Alternating$^\dagger$
    & \underline{2.35} & 6.92 & 12.41
    & 8.67 & 19.00 & 9.87
    & 10.27 & 13.67 & 36.62 & 4.73 \\
    Time / Alternating
    & 2.84 & 11.41 & 16.95
    & 13.21 & 24.00 & 13.68
    & 15.88 & 19.28 & 42.98 & 4.93 \\
    Time / Note memory
    & 2.37 & 7.50 & 12.45
    & \underline{8.33} & \underline{16.61} & 9.45
    & 11.00 & 14.40 & 28.16 & 4.41 \\
    Time / Frame memory
    & 2.58 & 7.93 & 12.91
    & 9.61 & 18.41 & 10.29
    & 11.53 & 14.94 & \underline{27.82} & 4.53 \\
    Time--Pitch / Note memory
    & 2.37 & 7.04 & 12.22
    & 8.35 & 16.89 & \underline{9.37}
    & 10.45 & 13.85 & \textbf{26.50} & \underline{4.11} \\
    Time--Pitch / Frame memory
    & 2.55 & 9.73 & 14.82
    & 11.18 & 20.54 & 11.76
    & 13.48 & 16.89 & 30.38 & 5.16 \\
    \cmidrule(lr){1-11}
    Mask 0 / 0
    & 2.53 & 6.44 & 11.62
    & 11.28 & 20.01 & 10.38
    & 10.57 & 13.98 & 35.42 & 5.73 \\
    Mask 10 / 0
    & 2.42 & 6.99 & 12.18
    & 12.07 & 21.69 & 11.07
    & 11.45 & 14.86 & 43.21 & 5.13 \\
    Mask 0 / 75
    & 2.43 & 7.08 & 12.31
    & 9.00 & 17.26 & 9.62
    & 10.70 & 14.11 & 30.87 & 4.29 \\
    \midrule
    \multicolumn{11}{c}{\textbf{Gangqinpu}} \\
    DuET
    & 0.63 & \underline{1.66} & 2.28
    & \textbf{1.75} & \textbf{9.01} & \textbf{3.07}
    & 2.26 & 1.78 & 35.26 & 2.72 \\
    Note encoder only
    & 0.62 & 1.77 & \underline{2.07}
    & 2.87 & 10.88 & 3.64
    & 2.43 & 1.95 & \textbf{22.93} & \textbf{2.38} \\
    \cmidrule(lr){1-11}
    Index / Alternating$^\dagger$
    & 0.64 & 1.94 & 3.08
    & 2.54 & 11.07 & 3.85
    & 2.36 & 1.88 & 54.11 & 2.91 \\
    Time / Alternating
    & 0.63 & 1.85 & 2.22
    & 2.89 & 10.28 & 3.57
    & 2.78 & 2.30 & \underline{26.52} & 2.95 \\
    Time / Note memory
    & \textbf{0.60} & \textbf{1.58} & \textbf{2.05}
    & \underline{1.76} & 9.36 & \underline{3.07}
    & \underline{2.24} & \underline{1.76} & 32.82 & 2.59 \\
    Time / Frame memory
    & 0.71 & 2.12 & 2.41
    & 2.93 & 9.63 & 3.56
    & 3.04 & 2.55 & 32.19 & \underline{2.46} \\
    Time--Pitch / Note memory
    & 0.60 & 1.71 & 2.18
    & 2.49 & 9.52 & 3.30
    & 2.38 & 1.90 & 35.44 & 2.66 \\
    Time--Pitch / Frame memory
    & 0.69 & 2.44 & 3.15
    & 3.44 & 10.19 & 3.98
    & 3.06 & 2.57 & 31.25 & 2.87 \\
    \cmidrule(lr){1-11}
    Mask 0 / 0
    & 0.64 & 1.83 & 2.59
    & 3.23 & 10.04 & 3.67
    & 2.57 & 2.09 & 37.65 & 2.78 \\
    Mask 10 / 0
    & 0.67 & 1.95 & 2.58
    & 3.30 & 9.64 & 3.63
    & 2.78 & 2.30 & 36.42 & 2.53 \\
    Mask 0 / 75
    & \underline{0.60} & 1.69 & 2.70
    & 1.87 & \underline{9.06} & 3.18
    & \textbf{2.23} & \textbf{1.74} & 56.35 & 2.67 \\
    \bottomrule
    \end{tabular*}

    \vspace{2pt}
    \begin{minipage}{\linewidth}
    \scriptsize
    Names follow ``RoPE / decoder memory'': Index, Time, and Time--Pitch denote the positional encoding. Alternating uses note memory in odd decoder layers and frame memory in even layers; Note memory or Frame memory uses that memory in all eight layers. These variants retain both encoders; Note encoder only removes the frame branch. DuET uses Time--Pitch / Alternating. $^\dagger$The index-based variant also receives inter-onset intervals. Mask rates are input-bar / decoder-history-note percentages
(10 / 75 for DuET).
    \end{minipage}
    \vspace{-12pt}
\end{table}

\begin{table}[t]
    \centering
    \fontsize{7.5}{8.5}\selectfont
    \setlength{\tabcolsep}{5pt}
    \renewcommand{\arraystretch}{0.92}
    \setlength{\aboverulesep}{1pt}
    \setlength{\belowrulesep}{1pt}
    \caption{Beat-tracking ablations on 59 ASAP and 100 Gangqinpu
    performances. Values are macro F1 scores (\%); higher is better.
    Bold and underlined values indicate the best and second-best results
    in each column. Positional/memory variants are named as
    ``RoPE / decoder memory''. DuET uses Time--Pitch RoPE with
    alternating memory access.}
    \label{tab:beat_ablations}
    \begin{tabular}{@{}lrrrr@{}}
    \toprule
    & \multicolumn{2}{c}{ASAP}
    & \multicolumn{2}{c}{Gangqinpu} \\
    \cmidrule(lr){2-3}\cmidrule(lr){4-5}
    Model & Beat & Downbeat & Beat & Downbeat \\
    \midrule
    DuET
    & 68.01 & \textbf{65.55} & 77.34 & 79.94 \\
    Frame-only tracker
    & 61.19 & 55.51 & 78.86 & \underline{81.37} \\
    Dual-encoder tracker
    & 62.89 & 53.97 & \underline{80.24} & 80.73 \\
    \cmidrule(lr){1-5}
    Index / Alternating$^\dagger$
    & 62.34 & 61.66 & 70.60 & 73.09 \\
    Time / Alternating
    & 65.73 & 60.45 & \textbf{81.24} & \textbf{82.50} \\
    Time / Note memory
    & 68.42 & 63.84 & 76.07 & 79.36 \\
    Time / Frame memory
    & 68.26 & \underline{64.42} & 79.26 & 80.58 \\
    Time--Pitch / Note memory
    & \underline{69.09} & 62.66 & 73.31 & 77.46 \\
    Time--Pitch / Frame memory
    & 65.60 & 62.09 & 78.57 & 80.78 \\
    \cmidrule(lr){1-5}
    Mask 0 / 0
    & 64.12 & 62.07 & 74.58 & 80.25 \\
    Mask 10 / 0
    & 67.02 & 62.81 & 77.14 & 78.25 \\
    Mask 0 / 75
    & \textbf{69.29} & 64.23 & 72.35 & 74.68 \\
    \bottomrule
    \end{tabular}

    \vspace{2pt}
    \begin{minipage}{\linewidth}
    \fontsize{7.5}{8.5}\selectfont
    The two dedicated trackers omit the score decoder.
    All positional/memory variants retain both encoders and an
    eight-layer decoder.
    $^\dagger$The index-based variant additionally receives inter-onset
    intervals as input features.
    Mask rates denote encoder-input / decoder-history percentages;
    DuET uses 10 / 75.
    \end{minipage}
\end{table}

\subsection{Ablations}
\label{sec:ablations}

We examine encoder branches, positional encoding, decoder memory access, and masking in \autoref{tab:pm2s_ablations} and \autoref{tab:beat_ablations}. PM2S results use fixed common-valid subsets of 39 ASAP and 72 Gangqinpu performances, retained from the original ablation suite. Beat tracking uses all 59 ASAP and 100 Gangqinpu test performances.

\paragraph{Encoder branches and joint supervision.}
Removing the frame branch increases MUSTER mean error from $8.68\%$ to $10.50\%$ on ASAP and from $3.07\%$ to $3.64\%$ on Gangqinpu. Onset and offset errors also increase on both datasets, supporting the value of frame-level temporal context. Duration results are mixed: DuET achieves lower ScoreSimilarity duration error on ASAP, whereas Note encoder only performs better on Gangqinpu. Among the dedicated trackers, adding the note branch improves beat F1 on both datasets but reduces downbeat F1. Full DuET achieves higher beat and downbeat F1 than both dedicated trackers on ASAP, whereas both trackers perform better on Gangqinpu. Tracker checkpoints are selected using beat metrics and score models using score metrics, so these differences cannot be attributed solely to joint supervision.

\paragraph{Positional encoding.}
Index / Alternating yields higher MUSTER mean error than DuET on both datasets, despite additionally receiving inter-onset intervals as input features. This suggests that performance-based positional encoding can improve transcription accuracy in the proposed architecture. With alternating memory access, Time--Pitch RoPE reduces MUSTER mean error relative to time-only RoPE from $13.68\%$ to $8.68\%$ on ASAP and from $3.57\%$ to $3.07\%$ on Gangqinpu. However, its effects under single-memory decoding are mixed: Time--Pitch RoPE improves mean error with note memory on ASAP, but increases it with note memory on Gangqinpu and with frame memory on both datasets. Thus, the benefit of pitch-aware encoding depends on decoder memory access and the dataset.

\paragraph{Decoder memory access.}
With Time--Pitch RoPE, alternating access to both memories achieves lower MUSTER mean error than either Note memory or Frame memory on both datasets. This supports allowing the decoder to access both note relationships and frame-level temporal context. With time-only RoPE, however, both single-memory variants achieve lower mean error than alternating access on both datasets. Several single-memory variants also obtain lower duration errors than DuET. These results indicate that the effectiveness of memory access depends on the positional encoding.

\paragraph{Masking.}
DuET uses encoder-input bar masking and decoder-history note masking at rates of $10\%$ and $75\%$, respectively (see Section~\ref{sec:masking}). Removing either masking mechanism increases MUSTER mean error on both datasets. Removing decoder-history masking also substantially increases onset error, consistent with its role in reducing reliance on preceding score attributes. The combined setting nevertheless does not minimize every metric: removing input masking improves Gangqinpu note-deletion and note-insertion errors, and some masking variants achieve higher beat F1.

\subsection{Qualitative analysis}
\label{sec:qualitative}

\begin{figure}[t]
    \centering
    \includegraphics[width=\linewidth]{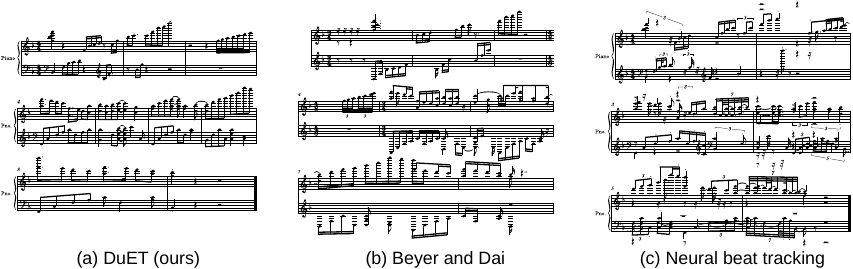}
    \caption{Qualitative comparison of score excerpts reconstructed from the same performance MIDI: (a) DuET, (b) Beyer and Dai, and (c) Liu's PM2S. All outputs are rendered with MuseScore 4.}
    \label{fig:qualitative}
\end{figure}

For qualitative analysis, we randomly selected an excerpt from a piano performance on YouTube\footnote{\url{https://www.youtube.com/watch?v=UkLJl8lEChY}} and transcribed it into performance MIDI using Transkun v2. \autoref{fig:qualitative} compares the scores reconstructed from this MIDI by DuET and two neural baselines. DuET produces a more regular rhythmic organization, with consistent $4/4$ meter and relatively simple note values. \citet{beyer2024end} predict changes between $3/4$ and $4/8$ and introduce triplet groupings, while the model of \citet{liu2022performance} produces frequent tuplets, fragmented rests, and overlapping voices. These examples illustrate how different interpretations of expressive performance timing can lead to substantially different score structures. DuET yields a more coherent and readable transcription in this excerpt, although visual simplicity alone does not establish transcription accuracy.

\section{Conclusion}
\label{sec:conclusion}

We presented DuET, a dual-encoder Transformer for piano PM2S transcription that combines note-level score prediction with frame-level metrical supervision. Bidirectional encoder cross-attention and alternating decoder cross-attention integrate note relationships with temporal context, while TimeRoPE and TPRoPE represent relative performance time and pitch intervals. Experiments on ASAP and Gangqinpu show consistent improvements in timing-related metrics, with a substantial reduction in overall transcription error on Gangqinpu. Ablations demonstrate the value of the frame branch and decoder-history masking, although the benefits of pitch-aware encoding and memory integration vary across datasets and metrics. Future work will improve downbeat transcription accuracy, and extend evaluation to broader musical styles.

\bibliographystyle{unsrtnat}
\bibliography{ref}

\end{document}